\documentclass{article}
\usepackage{xcolor}
\usepackage{frascatiphys}
\usepackage{graphicx}
\usepackage{cancel}
\usepackage{hyperref}

\begin{document}
%
\title{ Dark matter from WIMP to FIMP over three regimes: Cosmology versus
  Colliders } \author{Laura Lopez-Honorez$^{1,2}$, Quentin
  Decant$^{1}$, Sam Junius$^{1,2,3}$ and Michel H.G. Tytgat$^{1}$
  \\ {\em $^1$ Service de Physique Th\'eorique, Universit\'e Libre de
    Bruxelles, C.P. 225, 1050 Brussels, Belgium. } 
  \\
    {\em $^2$ Theoretische Natuurkunde \& The International Solvay Institutes,} 
     \\
   {\em $^3$ Inter-University Institute for High Energies, Vrije Universiteit Brussel,
      Pleinlaan 2, 1050 Brussels, Belgium.}
}
\maketitle
\baselineskip=10pt
\begin{abstract}
 Dark matter models can give rise to specific signatures at
 particle physics experiments or in cosmology. The details of the
 cosmological history can also influence the new physics signals to be
 expected at e.g. collider experiments. In these proceedings, we
 briefly summarize the case of dark matter weakly to feebly coupled to
 Standard Model fermions through  $t$-channel portal dividing the
 discussion into three main regimes. We also underline the interplay
 between cosmology and particle physics.
\end{abstract}
\baselineskip=14pt

\section{Introduction}

Cosmological observations imply that around 80\% of the total matter
content in our universe is made up of dark matter
(DM).\cite{Planck:2018vyg} Despite substantial effort, searches in
colliders, direct, and indirect experiments have so far not yielded
any clear hints of interactions other than gravitational between the
DM and the Standard Model (SM) particles. Weakly interacting massive
particles (WIMP), produced through dark matter annihilation or
co-annnihilation driven freeze-out (FO) still agree with such absence
of DM signal in, sometimes large, parts of their parameter space. On
the other hand, feebly interacting dark matter particles (FIMP), more
feebly coupled to the SM than WIMP, can easily evade WIMP
searches. Here we first briefly review the possible mechanisms for DM
production in the early universe, from WIMP to FIMP, going through DM
(co-)annihilation, mediator annihilation and conversion driven FO,
freeze-in (FI) and SuperWIMP (SW) mechanisms. We discuss then possible
probes of dark matter including signatures in particle physics
detectors and cosmology.

\begin{figure}
  \begin{center}
  \includegraphics[scale=0.35]{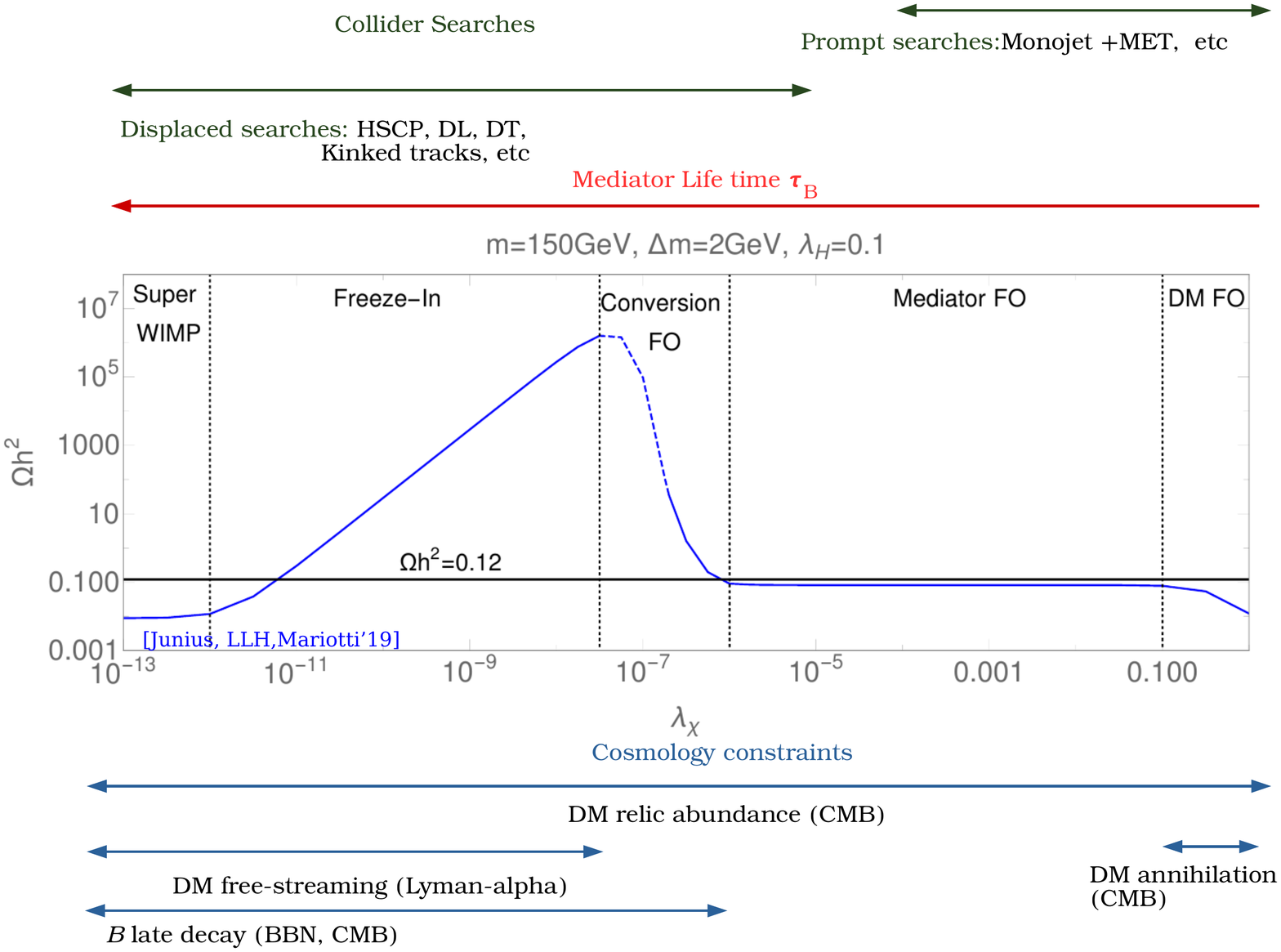}  
  \end{center}
  \caption{\it DM abundance and production mechanisms as a function of
    the coupling for a compressed DM-mediator mass spectrum in a
    leptophilic scenario, see\cite{Junius:2019dci} as well as\cite{Junius:2022th}. As a guide for the eye, we show with arrows, above and
    below the plot, the range of couplings that may be probed at colliders
    and with cosmology data.  }
  \label{fig:relic}
\end{figure}
For concreteness, we focus here on top-philic and leptophilic dark
matter, $\chi$, in the form of a real scalar singlet or a Majorana
fermion. We consider a DM coupling through a Yukawa type of
interaction to a right-handed SM fermion $f_R$, being a top $t_R$ or a
light lepton $\ell_R$, and a new charged dark sector bath particle,
$B$. The lagrangian of the interaction reads:
\begin{equation}
  {\cal L}\supset \lambda_\chi \Phi_B \bar \chi f_R+h.c. \quad {\rm or}
  \quad {\cal L}\supset \lambda_\chi \chi \bar \Psi_B f_R+h.c.
\end{equation}
where $B=\Phi_B (\Psi_B)$ denotes a charged scalar (fermion) bath
particle. This particle is always in thermal contact with the SM in
the early universe because of its gauge couplings and plays the role
of the DM-SM mediator of interaction. We also assume that both dark
sector particles $B$ and $\chi$ are odd under a $Z_2$ symmetry while
the SM particles are even. This ensures the DM stability. These type
of DM scenarios are sometimes referred to as $t$-channel DM or vector-like portal to DM in the case of a WIMP and a scalar DM candidate,
respectively. Let us emphasize that even though they correspond to
rather minimal extensions of the SM, their DM phenomenology is very
rich and has been investigated in multiple works.\footnote{We
apologize for the impossibility to cite all of the relevant works in
these proceedings given the page number constraints. A more extended
list of references on $t$-channel scenarios and the corresponding DM bounds from colliders, cosmology, etc can be found in the reference papers cited here. }
Figure~\ref{fig:relic} illustrates the interplay between cosmology and
colliders for a fermion leptophilic DM coupling to a muon as a
function of the coupling $\lambda_\chi$. The plot
from\cite{Junius:2019dci} shows the evolution of the DM relic
abundance $\Omega h^2$ as a function of $\lambda_\chi$ for a fixed
value of the DM mass $m_\chi=150$ GeV, and a small mass splitting of 2
GeV between the mediator $B$ and the DM. A qualitatively similar
result could be obtained for top-philic DM.  By varying $\lambda_\chi$
between $\sim {\cal O}(1)$ and $10^{-14}$, we go from WIMP production
through DM annihilation FO to FIMP production through SW. Above and
below the plot, the green and blue arrows indicate for which typical range
of $\lambda_\chi$ colliders and cosmology probes may test the DM
model. In what follows, we separate the discussion in three different regimes.

\section{WIMP from DM (co-)annihilation freeze-out }

Vanilla WIMP DM is in chemical and kinetic equilibrium with the SM
plasma in the early universe and decouples chemically, or freezes-out,
when the rate of DM annihilations becomes too slow compared to the
expansion rate of the universe. In the latter case one expects the DM
relic abundance to be inversely proportional to the DM annihilation
cross-section: $\Omega h^2\sim \langle \sigma
v\rangle_{\chi\chi}^{-1}$. This implies that $\Omega h^2$ should
decrease for increasing values of the coupling as visible in the right
corner of Fig.~\ref{fig:relic}. For the models considered here one
would thus expect that $\Omega h^2\sim \lambda_\chi^{-4}$ for
annihilations through $t$-channel exchanges of the mediator $B$ and a
pair of fermions in the final state. 
Also, the presence of the mediator
allows to account for the right DM abundance for small couplings
$\lambda_\chi< 0.1$ and compressed spectra. In the latter cases, the
DM relic abundance can become driven by co-annihilations (or mediator
annihilations) with $\Omega h^2\sim \langle \sigma
v\rangle_{B\chi}\sim g^2 \lambda_\chi^2 $ (or $\Omega h^2\sim \langle
\sigma v\rangle_{BB}\sim g^4 $), where $g$ denotes the gauge coupling
of the mediator $B$. This is well visible in Fig.~\ref{fig:relic}
around $\lambda_\chi\sim 0.1$ where the relic abundance becomes less
and less sensitive to $\lambda_\chi$ for decreasing values of the
latter. Naively, the main indirect dark
matter signatures would be expected to come from DM annihilations into
leptons or quarks. Let us emphasize though that, for Majorana or real
scalar dark matter, the annihilation into a pair of light fermions is either
$p$-wave or $d$-wave chirally suppressed and radiative processes, such
as virtual internal Bremssthralung or loop induced annihilations into
vector bosons, may leave the most promising signatures in indirect
searches through gamma ray lines or even drive the dark matter relic abundance in the case of
scalar DM, see e.g.\cite{Giacchino:2013bta}.

\begin{figure}[t]
    \begin{center}
        {\includegraphics[scale=0.35]{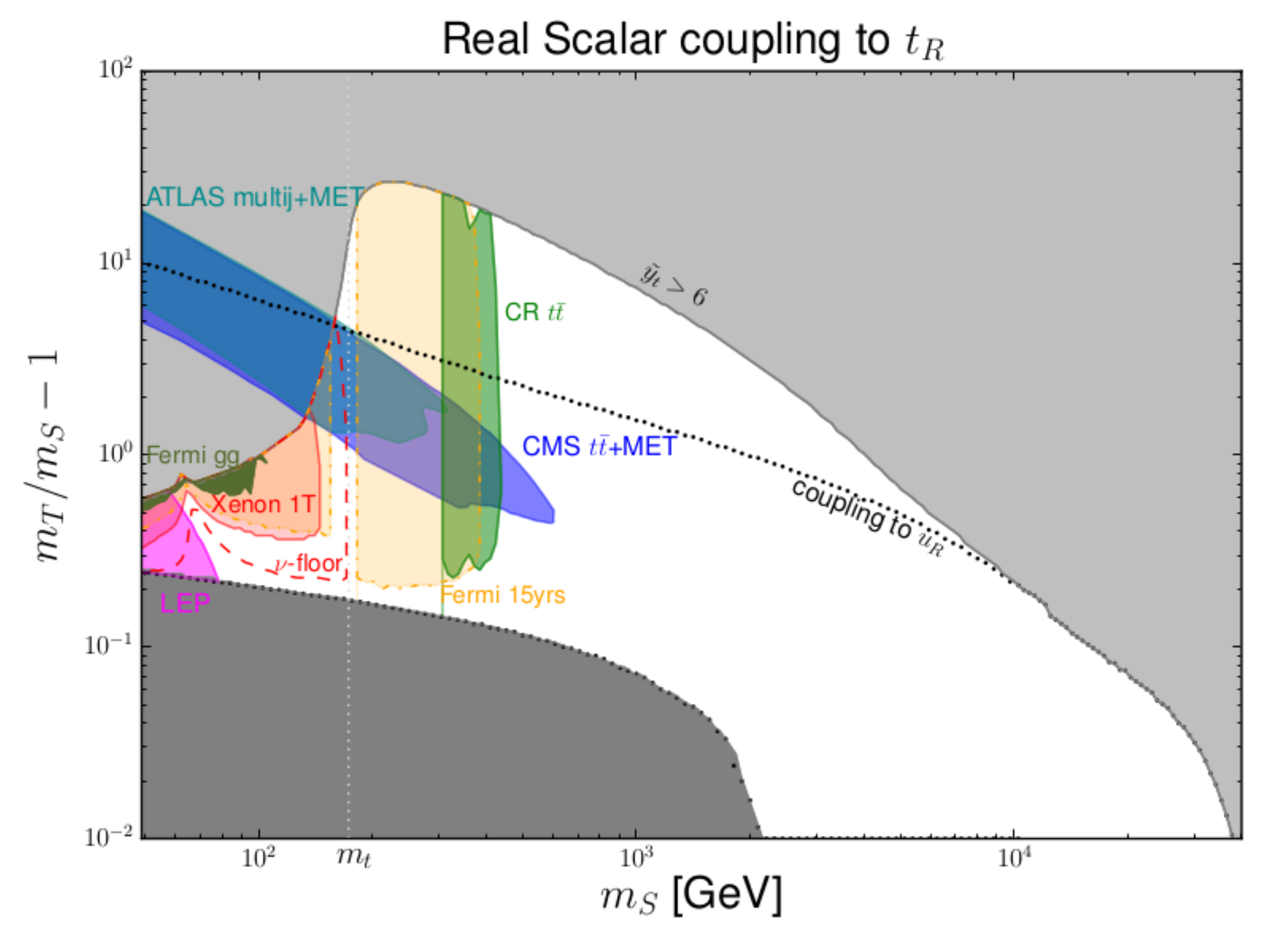}}\hspace{0.5cm}
        {\includegraphics[scale=0.45]{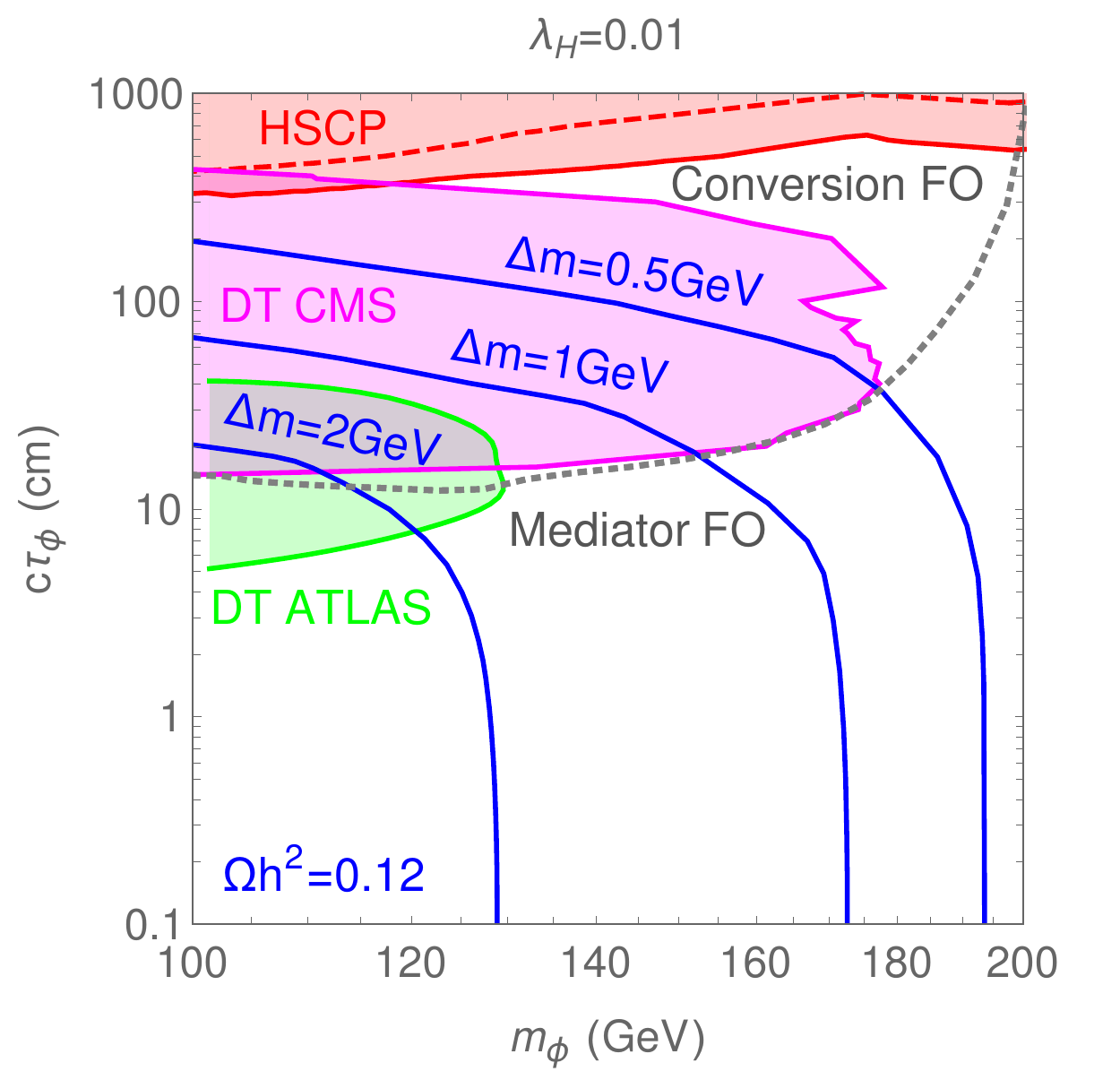}}
        \caption{\it Viable parameter space and constraints for top-philic scalar DM from DM (co-)annihilations driven
          freeze-out (Left)\cite{Colucci:2018vxz} and leptophilic fermionic DM coupling to
          $f_R=\mu_R$ from mediator annihilations and conversion driven
          freeze-out (Right)\cite{Junius:2019dci}.}
\label{WIMP}
    \end{center}
\end{figure}

In the left panel of Fig.~\ref{WIMP} we show the viable parameter
space of a top-philic scalar DM obtained in\cite{Colucci:2018vxz} in
the relative mass splitting $m_B/m_\chi-1$ versus dark matter mass
plane. In the white region, the right abundance is obtained for
couplings $\lambda_\chi\sim 10^{-2}-{\cal O}(1)$ through
(co-)annihilation FO while, for the lowest values of the relative mass
splitting, FO through mediator annihilations is giving the DM
abundance for even more suppressed values of $\lambda_\chi$. In the WIMP
top-philic scenario, collider searches for prompt signatures (ATLAS and
CMS in blue and LEP in pink) provide complementary constraints to
direct dark matter searches in red and indirect dark matter searches
in yellow and green. Still, a very large part of the viable parameter
space remains unconstrained, even when updated constraints from LHC
are taken into account. Notice that a detailed treatment of radiative
corrections to vector-like portal DM necessary for this analysis was
provided in\cite{Colucci:2018qml}. A review article on WIMP
$t$-channel fermion DM can be found in\cite{Garny:2015wea}. 
Also notice that it is actually possible to
get the right relic abundance of DM in the bottom dark gray zone
through conversion driven FO introduced in\cite{Garny:2017rxs}. This
is addressed in the next section.

\section{FIMP from DM mediator and conversion driven freeze-out }

The fact that the relic abundance can be driven by co-annihilations
and or, even more importantly, mediator annihilations is based on the
assumption that the rate of $B\leftrightarrow \chi$ conversions is
fast enough to keep DM in chemical equilibrium with the bath. For
leptophilic DM, in Fig.~\ref{fig:relic}, this assumption breaks when
the relic abundance begins to increase again for decreasing values of
$\lambda_\chi$ around $10^{-6}$. Notice that such a small coupling
implies the possibility to produce a long lived bath particle, coupled
through gauge interactions, at colliders and potentially give rise to
displaced signatures. In the case of mediator annihilation driven FO,
arising for $\lambda_\chi\sim 10^{-6}-10^{-2}$, the $B$ annihilations
and the relative mass splitting set the DM relic abundance. $\Omega
h^2$ shows no dependence in $\lambda_\chi$. When $\lambda_\chi<
10^{-6}$, $B\leftrightarrow \chi$ conversions are no longer fast
enough to convert DM back to bath particles efficiently and $n_\chi>
n_\chi^{\rm eq}$ before FO. The smaller the coupling, the slower are
the conversions and the larger is $n_\chi$ at FO. This is the reason
why $\Omega h^2$ increases with decreasing $\lambda_\chi$. This
mechanism for setting the FIMP relic abundance has been dubbed
conversion driven FO\cite{Garny:2017rxs}.

Here we show the viable parameter space for a leptophilic fermion DM
coupling to $\mu_R$ in the left panel of Fig.~\ref{WIMP},
see\cite{Junius:2019dci}, in the plane of the lifetime of the scalar
bath particle $B=\Phi_B$, denoted as $\tau_\phi$, as a function of its
mass, denoted as $m_\phi$. Going to the upper part of the plot we are
considering long lived $\Phi_B$, with decay lengths of few centimeters
or more. This corresponds to lower values of $\lambda_\chi$ and more
compressed spectra. The relic abundance is driven by conversions
(mediator annihilations) above (below) the gray dashed line. We
directly see that the full viable parameter space for such production
mechanism is already well constrained by displaced searches at
colliders looking for heavy stable charged particles (HSCP) and
disappearing tracks (DT), with the final leptons being too soft to be
detected.

\section{FIMP from freeze-in and superWIMPs }

As can be seen in Fig.~\ref{fig:relic}, another change in the relic
abundance behaviour happens for even smaller couplings around
$\lambda_\chi\sim$ few $\times 10^{-8}$. At that point, the DM can not
be expected to have ever been near kinetic or chemical equilibrium
with the bath. The relic abundance then decreases with decreasing
values of the coupling. This is the typical behaviour of DM production
from FI through $B$ decays. In the latter case, $\Omega h^2\propto
R_\Gamma$, where $R_\Gamma\sim M_p\Gamma_B/m^2_B$ is  directly
proportional to the bath particle decay rate with $\Gamma_B\propto
\lambda^2_\chi$, i.e. $\Omega h^2\propto \lambda_\chi^2$. The DM
abundance is then set when the bath particle becomes non-relativistic
and its number density becomes Boltzmann suppressed. In contrast, for
$\lambda_ \chi<10^{-12}$, the DM abundance becomes again independent of
$\lambda_ \chi$. This happens when the mediator lifetime is so long
that it essentially decays after $B$ chemical decoupling. In the
latter case, for the models under study, the relic abundance is set by
the so-called superWIMP mechanism and $\Omega_\chi=m_\chi/m_B\times
\Omega_B$, i.e. the final DM relic abundance is independent of
$\lambda_\chi$. In practice, a combination of FI and SW production can
also contribute to the DM abundance.

\begin{figure}[htb]
    \begin{center}
     {\includegraphics[scale=0.56]{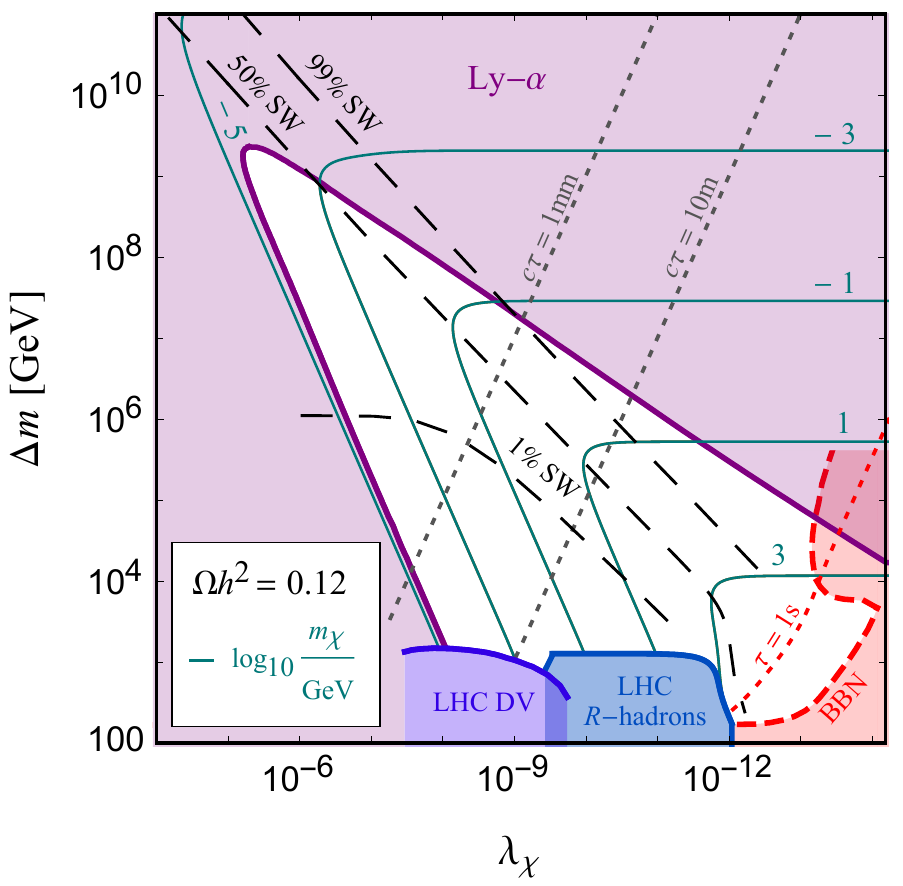}}\hspace{0.5cm}
       {\includegraphics[scale=0.52]{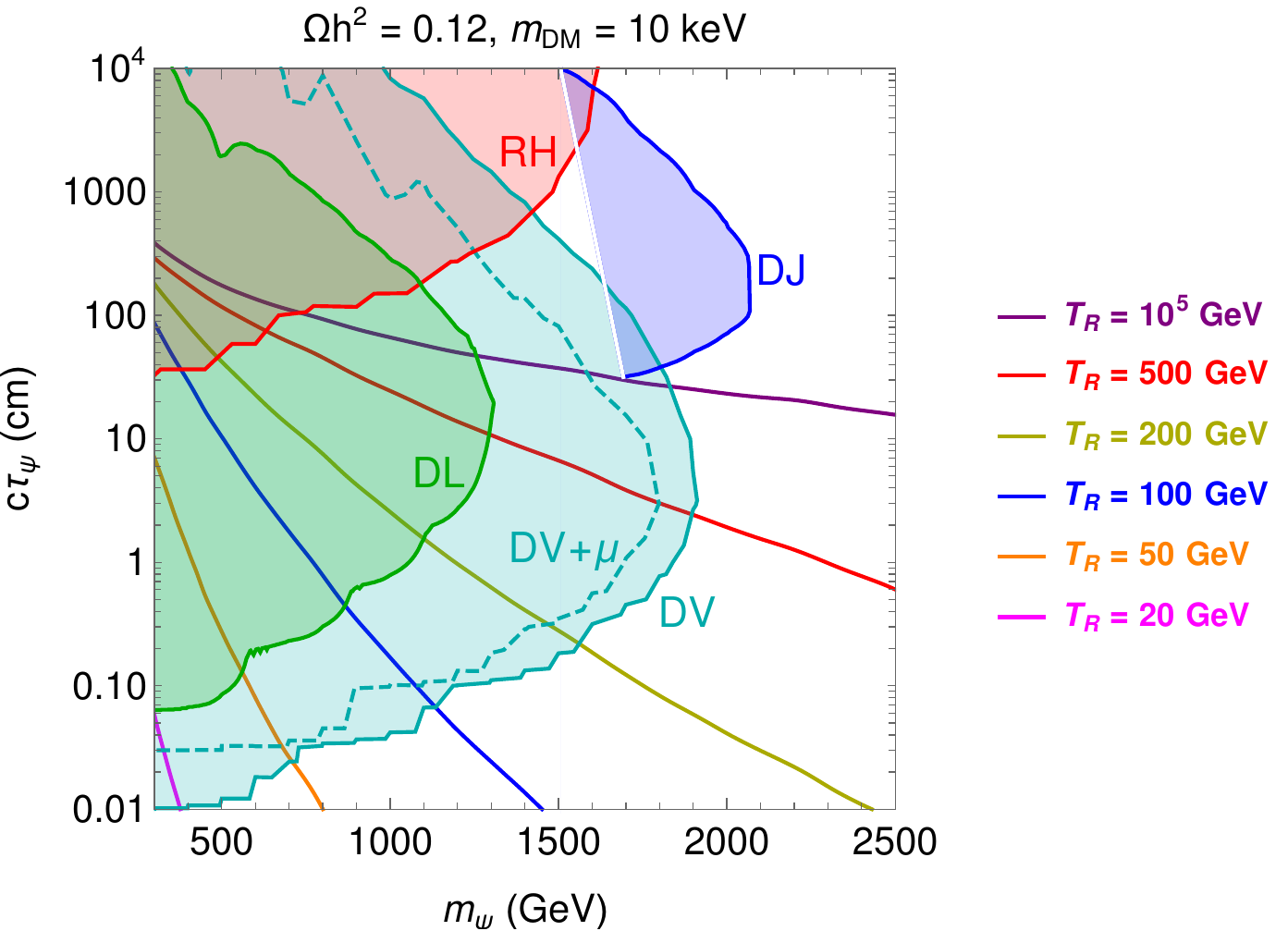}}
        \caption{\it FIMP viable parameter space and constraints for
          top-philic DM in the case of fermion DM from FI and SW
          (left)\cite{Decant:2021mhj} and of scalar DM from FI
          allowing for an early matter-dominated era (right)\cite{Calibbi:2021fld}.}
        \label{FIMP-FISW}
    \end{center}
\end{figure}

This is illustrated in the left panel of Fig.~\ref{FIMP-FISW}, where
we show the viable parameter space for top-philic fermionic dark
matter in the plane of $\Delta m= m_B-m_\chi$ as a function of
$\lambda_\chi$ from\cite{Decant:2021mhj}. The right FIMP relic
abundance is obtained along the cyan lines for different DM
masses. The dashed black lines inform on the relative contribution of
FI and SW to $\Omega_\chi h^2$. The colored areas are excluded by
cosmological constraints (magenta for Lyman-$\alpha$, and red for Big
Bang Nucleosynthesis) and collider constraints (from displaced
vertices, in purple, and R-hadrons searches, in blue
). In\cite{Decant:2021mhj} Lyman-$\alpha$ forest constraints on
thermal warm DM were carefully re-interpreted to put generic
constraints on free-streaming FIMPs excluding $m_\chi<15$ keV for FI
and masses as large as few GeV for SW. Very interestingly the left
panel of Fig.~\ref{FIMP-FISW} shows the nice interplay between
collider and cosmology to corner the viable parameter space
of FIMPs.

Now all the above results have been obtained assuming a standard
cosmological history. In particular, it was assumed that the DM was
produced in a radiation dominated era. When considering an early
matter dominated era, with e.g. a low reheating temperature $T_R$, the
relic abundance of dark matter can be diluted due to late time entropy
injection. Interestingly for FI, this implies larger values of
$\lambda_\chi$ or equivalently shorter $B$ lifetimes/decay
lengths. As a result, a larger part of the parameter space of FI
scenarios can be tested at colliders through displaced searches. This
is illustrated in the right panel of Fig.~\ref{FIMP-FISW}
from\cite{Calibbi:2021fld} in the case of a top-philic scalar DM
scenario in the plane of the lifetime of the fermionic bath
particle as a function of its mass, denoted with $\tau_\Psi$ and
$m_\Psi$,  respectively. The colored areas are excluded by
displaced vertices +MET (DV) in light blue, delayed jet (dark blue)
and R-hadron searches (red). The continuous colored lines serve as a
rough estimate of the Lyman-$\alpha$ constraints for reheating
temperatures between 20 GeV and $10^5$ GeV. The surface below these
lines would be excluded. As visible from the plot, with low $T_R= 20$
GeV displaced searches are much more efficient to exclude and probe in
the future the small decay length parameter space. This illustrates
yet another interesting interplay between cosmology and collider
experiments. Also notice that in\cite{Calibbi:2021fld} a
classification of DM scenarios for three-body interactions
between DM, $B$ and SM particles was proposed and a systematic
analysis of prospects for detection at colliders through displaced
signatures is provided.

\section{Conclusions}

Despite substantial experimental efforts dedicated to the search for
DM, no indisputable signature of DM has been found in (astro-)particle
physics experiments. In these proceedings we show that, even in a very
simple $t$-channel portal DM set up, a plethora of production
mechanisms are possible and a variety of DM probes are necessary to
pin point correctly the nature of dark matter. In the case of FIMPs in
particular, we underline the very nice interplay that exists between
cosmological probes, cosmological history and collider physics. It is
also worth mentioning that in all cases, from WIMP to FIMP, a
sometimes  large part of the parameter space is still viable and
potentially testable through specific signatures in colliders (with
e.g. displaced signals), astro-particle physics experiments (with line
like gamma-ray signals) and cosmology (with unconventional small-scale
structure evolution). It is only combining all possible probes of DM
that we can hope to conclusively test the DM nature.

\section{Acknowledgements}
We would like to thank all our collaborators for the work presented in
these proceedings and LLH thank the organisers of the LFC22 meeting
for the very nice workshop.  LLH and QD have been supported by the
Fonds de la Recherche Scientifique F.R.S.-FNRS through a research
associate position and a FRIA PhD grant.  LLH acknowledges support of
the ARC program of the Federation Wallonie-Bruxelles. LLH, QD and SJ
have been supported by the FNRS research grant number F.4520.19.  SJ
has also been supported by ULB and HEP\@ VUB. The work of MT  is
supported by the F.R.S.-FNRS under the Excellence of Science (EoS)
project No. 30820817 - be.h “The H boson gateway to physics beyond the
Standard Model”. All co-authors acknowledge support of the IISN
convention No. 4.4503.15.


%

%
\end{document}